\documentclass[letter]{aa} 

\usepackage{natbib,twoopt}
\usepackage[breaklinks=true]{hyperref} 
\bibpunct{(}{)}{;}{a}{}{,}             
\makeatletter
  \newcommandtwoopt{\citeads}[3][][]{\href{http://adsabs.harvard.edu/abs/#3}%
    {\def\hyper@linkstart##1##2{}%
     \let\hyper@linkend\@empty\citealp[#1][#2]{#3}}}
  \newcommandtwoopt{\citepads}[3][][]{\href{http://adsabs.harvard.edu/abs/#3}%
    {\def\hyper@linkstart##1##2{}%
     \let\hyper@linkend\@empty\citep[#1][#2]{#3}}}
  \newcommandtwoopt{\citetads}[3][][]{\href{http://adsabs.harvard.edu/abs/#3}%
    {\def\hyper@linkstart##1##2{}%
     \let\hyper@linkend\@empty\citet[#1][#2]{#3}}}
  \newcommandtwoopt{\citeyearads}[3][][]%
    {\href{http://adsabs.harvard.edu/abs/#3}
    {\def\hyper@linkstart##1##2{}%
     \let\hyper@linkend\@empty\citeyear[#1][#2]{#3}}}
\makeatother
\usepackage{graphicx}
\usepackage{txfonts}

\usepackage{algorithm}
\usepackage{algorithmic}

\usepackage{lscape}

\begin{document}

   \title{A machine learning framework to generate star cluster realisations}


   \author{George P. Prodan\fnmsep\thanks{prodangp9@gmail.com}
          \inst{1, 8}
          \and
          Mario Pasquato\inst{2, 3}
          \and 
          Giuliano Iorio\inst{1,4,5}
          \and
          Alessandro Ballone\inst{1,4,5}
          \and
          Stefano Torniamenti\inst{1,4, 6}
          \and 
          Ugo Niccolò Di Carlo \inst{7}
          \and
          Michela Mapelli\inst{1,4,6}
          }

   \institute{Dipartimento di Fisica e Astronomia, Università di Padova,
  Vicolo dell'Osservatorio 3, 35122, Padova, Italy 
         \and
             IASF Milano, via Alfonso Corti 12, Milano, Italy
        \and Ciela - Montreal Institute for Astrophysical Data Analysis and Machine Learning, Montréal, Canada 
        \and INFN–Padova, Via Marzolo 8, 35131, Padova, Italy 
        \and  INAF, Osservatorio Astronomico di Padova, Vicolo dell’Osservatorio 5, Padova, Italy 
        \and Institut f\"ur Theoretische Astrophysik, ZAH, Universit\"at Heidelberg, Albert-Ueberle-Str. 2, 69120, Heidelberg
        \and SISSA - Scuola Internazionale Superiore di Studi Avanzati, via Bonomea 365, I-34136 Trieste, Italy
        \and Faculty of Sciences, University of Craiova, A.I. Cuza 13, 200585 Craiova, Romania
             }

   \date{Received June 05, 2024
   ; accepted September 04, 2024}


  \abstract
   {Computational astronomy has reached the stage where running a gravitational $N$-body simulation of a stellar system, such as a Milky Way star cluster, is computationally feasible, but a major limiting factor that remains is the ability to set up physically realistic initial conditions.}
   {We aim to obtain realistic initial conditions for $N$-body simulations by taking advantage of machine learning, with emphasis on reproducing small-scale interstellar distance distributions.}
   {The computational bottleneck for obtaining such distance distributions is the hydrodynamics of star formation, which ultimately determine the features of the stars, including positions, velocities, and masses. To mitigate this issue, we introduce a new method for sampling physically realistic initial conditions from a limited set of simulations using Gaussian processes.}
   {We evaluated the resulting sets of initial conditions based on whether they meet tests for physical realism. We find that direct sampling based on the learned distribution of the star features fails to reproduce binary systems. Consequently, we show that physics-informed sampling algorithms solve this issue, as they are capable of generating realisations closer to reality.}
   {}


   \maketitle
%

\section{Introduction}

The target of humanity's first deliberate attempt at interstellar radio communication was the star cluster M13 \citep[][]{1975Icar...26..462.}, and this act is a testament to the importance of star clusters to astronomy, which can hardly be overstated. For instance, star clusters are the likely birthplace of most stars \citep[][]{2003ARA&A..41...57L}.

Astronomers use gravitational $N$-body simulations with the purpose of studying the evolution of star clusters \citep[][]{2003gnbs.book.....A, 2023LRCA....9....3S}. The goals, among others, are to predict gravitational wave emission from black holes and neutron stars formed by cluster stars \citep[][]{2021MNRAS.507.3612R, 2021MNRAS.508.3045D, 2023MNRAS.524..426I}, constrain the place of origin of the Solar System \citep[][]{2012AJ....143...73P, 2015PhyS...90f8001P}, determine the stability and habitability of exoplanets \citep[][]{2009ApJ...697..458S, 2011MNRAS.411..859M, 2012MNRAS.419.2448P}, and understand how star clusters contribute to the overall evolution of the Milky Way as part of the field called Galactic archaeology \citep[][]{2019ApJ...883L..31C, 2021ApJ...923...20P, 2022ApJ...931..156P}. 

Star clusters are understood to form within molecular clouds. Molecular clouds are high-density, low-temperature regions in the interstellar medium predominantly composed of molecular hydrogen. Star formation in molecular clouds occurs in small regions where the density becomes high enough (e.g. by turbulent colliding flows) to start a fast runaway gravitational collapse \citep{1999ARA&A..37..311E}. Gas in molecular clouds is often animated by turbulent motions, which are ultimately responsible for the irregular and complex spatial distribution of young cluster stars \citep{doi:10.1146/annurev-astro-091918-104430, Krause_2020, 2020MNRAS.496...49B}.
The scientific usefulness of gravitational $N$-body simulations depends on the realism of initial conditions (primordial positions, velocities, and masses of stars), which hinges on the ability to properly model star formation in molecular clouds. Realistic initial conditions may be obtained by running hydrodynamical simulations. These are computationally expensive, with one simulation taking around $10^5$ CPU hours \citep[][]{2021MNRAS.501.2920B}. 

Machine learning can be leveraged to obtain samples of valid initial conditions at a fraction of the computational price by learning a generative model from the outputs of a limited number of simulations. The first attempt in this direction was presented by \citet{Torniamenti_2021}. Their work consists of a bespoke manipulation of hierarchical clustering models that is simple and effective but lacks theoretical guarantees. Here we propose a different approach based on Gaussian processes (GPs).

\section{Gaussian processes}
Gaussian processes serve as probabilistic models capable of predicting function values based on noisy observations \citep{wang2022intuitive}. Especially valuable when data is limited, GPs present a more fitting alternative to deep neural networks given the constraints of our application \citep{https://doi.org/10.17863/cam.93643}. For star cluster realisations, the distribution of stars in a cluster can be represented as a function of key physical parameters, such as star masses, coordinates, and velocities. Leveraging GPs, we can learn the star distribution in clusters in the space defined by these parameters, enabling the generation of synthetic clusters. 

The essence of GP modelling lies in Bayesian inference, where model beliefs are updated upon receiving new observations. A GP model is characterised by a mean function, $\mu(x)$ and a kernel function, $K(x,x')$, with the kernel function reflecting the similarity degree between input points and influencing predictions \citep{Rasmussen2004}. In our case, the kernel function aims to represent the spatial relations among stars in a cluster by capturing the relationships among the stars set at star formation and arising through dynamics via, for example, the effects of gravitational attraction between the individual stars. 

The kernel function hyperparameters of GP models include the signal variance, $\sigma_y$; the length scale, $l$; and the noise level, $\sigma_n$. The signal variance measures signal amplitudes, the length scale indicates covariance decay distance, and the noise caters to errors. All of these parameters are trainable. Usually, the radial basis function kernel (RBF) is employed to compute the amplitude of the covariance function,
\begin{equation}
    \label{eq:rbf}
    K(x,x')=\sigma_y^2 \exp{\left[ {-\frac{(x-x')^2}{2l^2}} \right]}.
\end{equation}

The hyperparameters of the mean function can be defined depending on the choice of the prior mean function. For instance, a constant mean function $\mu(x)=c$ leads to one additional trainable hyperparameter.

Training aims to minimize the log marginal likelihood \citep{Rasmussen2004} with respect to the above-mentioned hyperparatemers $\theta=\{\sigma_y,\sigma_n, l, ...\}$, namely
\begin{align}
    \log P(y \ | \ X, \theta)=&-\frac{1}{2}(\mathbf{y-\mu})^T(K+\sigma_n^2I)^{-1}(\mathbf{y-\mu}) \notag \\
    &- \frac{1}{2}\log |K+\sigma_n^2I|-\frac{n}{2}\log 2\pi.
\end{align}
The first term ensures data fitting, the second one is responsible for regularisation, and the last one is a constant. Regularisation is an important part of model training that prevents overfitting (where the model learns irrelevant patterns in the data instead of the underlying relationships). By incorporating regularisation terms, the model generalizes better on unseen data, thus improving its predictive performance. Training with noise is an essential part of this work. By inserting noise, we obtained new cluster realisations that at the same time preserve some of the features of the training samples. Our aim for the GP fitting is to create a model that can reproduce these features.  

We drew the predictions from the posterior GP,
\begin{equation}
   y \sim \mathcal{GP}(\mu, K+\sigma_n^2 I), 
\end{equation}
where $\mu$, $K$ are the posterior mean and kernel function and $I$ is the identity matrix. One can draw two kinds of predictions. The first is the mean prediction, which generates the expected values that are similar to those determined for the training samples. This prediction is deterministic and represents the noiseless prediction of the GP. The second involves sampling from the GP's posterior distribution, incorporating noise. Here, the predictions are more diverse because they consider the inherent uncertainty in the model. This can lead to a broader range of potential outcomes and possibly to the formation of different clusters.
A complete mathematical overview on GP models is presented in \cite{books/lib/RasmussenW06}.

\section{Learning framework}
We have introduced a learning framework that utilizes GP modelling to learn the feature space distributions of stellar clusters, followed by sampling from these distributions to produce new clusters. This is an inverse problem, as the framework obtains new cluster realisations via simulation-based inference~\citep{doi:10.1073/pnas.1912789117, lueckmann2021benchmarking}.
While training, the inputs of the GP model are the parameters of each star, $\mathbf{\theta_i}^{(train)}$ with $i=\overline{1,N}$ for a cluster of $N$ stars. The parameters are physical quantities of the stars, with each star having a specific mass, $M$; location, $\mathbf{r}=(x,y,z)$; and velocity, $\mathbf{v}=(v_x,v_y,v_z)$, with respect to the cluster's centre of mass and all of the parameters $M$, $\mathbf{r}$, and $\mathbf{v}$ are not normalised. The GP learns a distribution over the probability density function of the feature space, $f(\theta)$.

Once the GP model input features were determined, our next step involved computing the target distribution, $y=f(\theta)$. For this purpose, $k$-nearest neighbours density estimators~\citep{knn} were implemented to ascertain the probability density values, $y_i$, which are normalised accordingly,
\begin{equation}
    y^{(train)}_i=\frac{Ny_i}{\Sigma_{j=1}^N y_j}.
\end{equation}

Further, we implemented Markov chain Monte Carlo (MCMC) approaches to simulate new cluster realisations based on the Metropolis-Hastings algorithm \citep{97f6cff1-d967-3115-9345-0aefb4d77f55, Ulam, Robert_2011}. The states proposed in the Markov Chain include a set of $N$ features at a given iteration step. We considered an iteration step $n$ corresponding to a state $S_n=\{\theta_1^{(n)}, \theta_2^{(n)}, ..., \theta_N^{(n)}\}$. The next candidate state, $S_{n+1}$, was obtained by a jump with probability $T(S_n \rightarrow S_{n+1})$. The acceptance probability of this jump is given by
\begin{equation}
 \label{eq:acc}
    \alpha=\min{\left(1, \frac{\pi(S_{n+1})T(S_n \rightarrow S_{n+1})}{\pi{(S_n)T(S_{n+1} \rightarrow S_{n})}}\right)},
\end{equation}
where $\pi$ is the stationary distribution of the chain. The probability of the new state can be written as the joint probability of the candidate features,
\begin{equation}
\label{eq:joint}
\pi(S_{n})=\prod_{i=1}^N f(\theta_i),
\end{equation}
where $f$ is a probability density function drawn from the GP model that is already trained on the features $\mathbf{\theta}^{(train)}$ and the corresponding target values $\mathbf{y^{(train)}}$ of the density function.

This sequential method first crafts a GP statistical model based on the probability density function of the features space for the cluster stars. While sampling from a standard probability density function is generally straightforward, our specific problem requires meticulous adjustments due to the need for maintaining physical constraints and the presence of complex correlations. This ensures that the generated samples accurately reflect the underlying distribution and maintains the validity of the initial conditions. We employed traditional MCMC methods based on the Metropolis algorithm to directly sample in a seven-dimensional space (DMCMC).
Moreover, we propose a physics-informed sampling approach based on learning the energy space distribution of nearest-neighbour star pairs (EMCMC). 

\subsection{Direct sampling}
\label{subsection:direct_sampling}
The direct approach is to define a seven-dimensional feature space that includes all the physical quantities of the stars in the clusters, meaning that for every star at the iteration of state $n$, we have $\theta^{(n)}_i=\{M_i^{(n)}, x_i^{(n)},y_i^{(n)},z_i^{(n)},v_{xi}^{(n)},v_{yi}^{(n)},v_{zi}^{(n)}\}$ with $i$ in $\overline{1,N}$. 
Considering the large number of stars in the cluster, achieving convergence becomes challenging if all parameters are subject to change each time we propose a new candidate cluster. However, Eq.~\ref{eq:acc} remains valid also if we propose only a new candidate star $\theta^{(n+1)}_i$ such that the new state becomes $S_{n+1}$ with 
\begin{equation}
    \theta^{(n+1)}_j=(1-\delta_{ij})\theta^{(n)}_i+\delta_{ij}\theta^{(n+1)}_i
\end{equation}
for $j$ in $\overline{1,N}$, where $\delta_{ij}$ is the Kronecker delta, and it is equal to one only if $i=j$; otherwise it is zero. The acceptance rate will be 
\begin{equation}
\label{eq:acc_new}
    \alpha=\min\left(1, \frac{f\left(\theta^{(n+1)}_i\right)}{f\left(\theta^{(n)}_i\right)}\right).
\end{equation}

Therefore, at each iteration a new subset of seven features $\theta_i^{(n+1)}$ is proposed with $i$ chosen randomly. We drew new candidates using a normal distribution such that $\theta_{ik}^{(n+1)} \sim \mathcal{N}(\theta_{ik}^{(n)}, \sigma)$ for every feature $k$ of $\theta_i$, where $\sigma$ is called the step size, and it is the amplitude of the perturbation applied on the previous state of the Markov chain and, mathematically, the standard deviation of the normal distribution centred on the previous sampled feature from which we sampled the new candidate.

The new candidate is accepted or rejected based on the rate established by Eq.~\ref{eq:acc_new} and employing the same exact density function $f$ drawn from the GP model, meaning that the 
GP noise seed must be kept unchanged during the sampling of one cluster realisation. The last condition is that it is essential not to break the detailed balance of the Metropolis-Hastings algorithm. 

One can notice that there is no conditioning on the global properties of the cluster and the sampler `builds' the cluster by sampling stars individually. Therefore, this direct approach works similar to a black box that relies solely on the correlations captured by the GP model.

\subsection{Physics-informed sampling}

Sampling star systems, especially binary systems in close interaction, by generating new samples in a seven-dimensional space is challenging due to the low probability of proposing candidate stars in proximity within the physical space. This may result in erroneous modelling of star clusters at small scales, even if the sampling algorithm successfully replicates the density function given by the GP. Our solution to this problem is based on incorporating physical laws to the sampling process by focusing on binary stars creating chains of pairs based on the nearest neighbours and learning the distribution over the potential and kinetic energies of these pairs. Thus, the cluster is reduced to a chain of this kind, and from this chain we can reconstruct other clusters based on similar chains.

This approach substitutes learning the full distribution of the mutual potential energies that results from $N(N-1)/2$ pairs of stars. This can be categorised as feature dimensionality reduction through feature selection \citep{jia_2022} based on the heuristic that most of the binding energy is concentrated in the binary systems of the cluster \citep{2021MNRAS.507.2253T}.

We defined for each pair a feature set based on the mutual potential energy of the stars and their kinetic energies, $\theta_i =(U_{i,i+1},K_i,K_{i+1})$ with $i$ in $\overline{1,N-1}$. In this way, following the nearest neighbours chain, we get $N-1$ sets of features in the energy space. We considered that these sets of energy values fully define the energy state of the nearest neighbours chain. New energy states are sampled making use of a GP model trained on the probability density function of the energy space.

The sampling scheme therefore starts by proposing new energy states. The first step of the solution is the EMCMC algorithm that samples the energy state of the nearest neighbours chain. The energy state at the $n$-th iteration, $S_n=\{\theta_1^{(n)}, \theta_2^{(n)}, ..., \theta_N^{(n)}\}$, implies the mutual potentials of the nearest neighbours, $U_{12}^{(n)}$, $U_{23}^{(n)}$, ..., $U_{N-1, N}^{(n)}$, and all the kinetic energies, $K_1^{(n)}$, $K_2^{(n)}$, ..., $K_N^{(n)}$. As before, we sampled for a random $i$ a new candidate $\theta_i^{(n+1)} =(U_{i,i+1}^{(n+1)},K_i^{(n+1)},K_{i+1}^{(n+1)})$ from a normal distribution based on the values of the previous iteration. This also implies a change for $\theta_{i-1}^{(n)}$ and $\theta_{i+1}^{(n)}$. In this case, applying Eq.~\ref{eq:acc} and Eq.~\ref{eq:joint} leads to the following acceptance rate:
\begin{equation}
    \label{eq:acc_emcmc}
        \alpha=\min\left(1, \frac{f\left(\theta^{(n+1)}_{i-1}\right)f\left(\theta^{(n+1)}_i\right)f\left(\theta^{(n+1)}_{i+1}\right)}{f\left(\theta^{(n)}_{i-1}\right)f\left(\theta^{(n)}_i\right)f\left(\theta^{(n)}_{i+1}\right)}\right),
\end{equation}
where $f$ is the probability density function drawn from the GP model trained on the energy space. We note that at the boundaries, only two sets of features are subject to change.

Using the new sampled chain, one needs to `reconstruct' the cluster by defining its stars as entities with a position, velocity, and mass. The magnitudes of the relative distances between the stars, $\mathbf{r_{i,i+1}}=\mathbf{r_{i+1}}-\mathbf{r_{i}}$, and the velocities are derived according to the physical laws of Newtonian gravity,
\begin{equation}
    |\mathbf{r_{i,i+1}}|=-\frac{GM_iM_{i+1}}{U_{i,i+1}}
    \label{eq:mj}
\end{equation}
\begin{equation}
    |\mathbf{v_{i+1}}|=\sqrt{\frac{2K_{i+1}}{M_{i+1}}},
    \label{eq:rv}
\end{equation}
where $G$ is the universal gravitational constant and $M_i$ and $M_{i+1}$ are the masses.

Following the chain sequentially implies that the properties of the star $i$ are known when deriving those of the next star in the chain (the star $i+1$). The mass of the next star, $M_{i+1}$, is sampled independently with the help of another GP model trained only on mass distributions. 
One can then determine the magnitudes of $\mathbf{r_{i,i+1}}$ and $\mathbf{v_{i+1}}$. The next step is to explore the full seven-dimensional feature space of the stars to determine the optimal directions of $\mathbf{r_{i,i+1}}$ and $\mathbf{v_{i+1}}$. For this, we used a GP model trained on the phase-space, $\mathcal{G}\mathcal{P}(\vec{r},\vec{v})$. The exploration consists of proposing several $N_c$ candidate directions for each vector. The candidates were chosen randomly from a uniform distribution. Using the GP model predictions, we chose the most probable position in the phase-space to determine the next star $i+1$. The procedure was repeated until the end of the chain.

In Algorithm \ref{alg:findnewstar}, we provide the pseudocode of the proposed algorithm that works based on three GP models: $\mathcal{G}\mathcal{P}(\theta)$, $\mathcal{G}\mathcal{P}(M)$, and $\mathcal{G}\mathcal{P}(\vec{r},\vec{v})$. The function `randomDirection' is employed to generate a random direction based on two random generated numbers, $\mathcal{R}_\theta$ and $\mathcal{R}_\phi$, which are used to generate pairs of angles $(\theta, \phi)$ that correspond to directions sampled uniformly on a three dimensional sphere. These directions are used to generate new candidates.

We define two hyper-parameters, $maxJump$ and $N_c$. The first parameter is used to control the maximum distance between the pairs of stars. In this way, the mass choice is conditioned on the spatial distribution of the cluster. Without this conditioning, the chain can be easily broken into multiple parts. This would lead to generating several smaller clusters. The other parameter, $N_c$, is the number of star candidates we generate randomly based on the values of $|\vec{r}_{ij}|$ and $|\vec{v}_{j}|$. We use $\mathcal{G}\mathcal{P}(\vec{r},\vec{v})$ afterwards to find the best candidate, that is, the star with the most probable position in the phase-space. As $N_c$ is large enough, the resulting star features always point towards the densest region of the phase-space that is attainable at a given moment.

\begin{algorithm}
\begin{algorithmic}
\REQUIRE $\mathcal{G}\mathcal{P}(\theta)$, $\mathcal{G}\mathcal{P}(M)$ , $\mathcal{G}\mathcal{P}(\vec{r},\vec{v})$ , \texttt{randomDirection}, $N$, $N_c$, $maxJump$
\STATE $cluster \gets \text{empty array to store the parameters of $N$ stars}$
\STATE $cluster[0] \gets \text{initialize parameters of the first star}$
\FOR {$i \texttt{ from 0 to N-1}$}
\STATE $U_{i,i+1}, K_{i}, K_{i+1} \gets MCMC(\mathcal{G}\mathcal{P}_{(\theta)})$

\STATE $M_i, \vec{r}_i, \vec{v}_i \gets cluster[i]$
\STATE $|\vec{r}_{i,i+1}| \gets \infty$
\WHILE{$|\vec{r}_{i,i+1}| > maxJump$}
\STATE $M_{i+1} \gets  MCMC(\mathcal{G}\mathcal{P}_{(M)})$
\STATE $|\vec{r}_{i,i+1}| \gets \frac{M_im_{i+1}}{U_{i,i+1}}$
\ENDWHILE

\STATE $|\vec{v}_{i+1}| \gets \sqrt{\frac{2 K_{i+1}}{m_{i+1}}}$
\STATE $candidates \gets \text{empty array to store $N_c$ candidates}$
\FOR{\texttt{each $candidate$ in $candidates$}}
    \STATE $\vec{\lambda}_{\vec{r}} \gets \texttt{randomDirection()}$
    \STATE $\vec{\lambda}_{\vec{v}} \gets \texttt{randomDirection()}$
    \STATE $\vec{r}_{i+1} = \vec{r}_i + |\vec{r}_{i,i+1}|\vec{\lambda}_{\vec{r}}$
    \STATE $\vec{v}_{i+1} = |\vec{v}_{i+1}|\vec{\lambda}_{\vec{v}}$
    \STATE $candidate \gets M_{i+1}, \vec{r}_{i+1}, \vec{v}_{i+1}$
\ENDFOR
\STATE $probs \gets \mathcal{G}\mathcal{P}_{(\vec{r},\vec{v})}(candidates)$
\STATE $bestCandidateIdx \gets argmax(probs)$
\STATE $cluster[i+1] \gets candidates[bestCandidateIdx]$
\ENDFOR
\end{algorithmic}
\caption{EMCMC algorithm in pseudocode to find and select new star candidates for the generated cluster.}
\label{alg:findnewstar}
\end{algorithm}

\section{Results and discussions}
\label{results}

\subsection{Dataset}
We utilized a dataset consisting of ten clusters extracted from hydrodynamical simulations conducted by \cite{Ballone_2020}. These simulations were designed to accurately reproduce not only the clumpiness but also the fractal nature observed in star-forming regions. Each cluster in the dataset has approximately $2500$ to $4000$ stars, with a cumulative mass ranging between $4000$ and $40000 \  M_\odot$ (solar masses). For our analysis, we divided this dataset into a training subset, consisting of seven clusters, and a validation subset made up of the remaining three clusters.

\subsection{Model training}
We used the gpytorch library \citep{gpytorch} to implement our models. This library offers a modular and efficient framework for GPs and supports graphics processing unit (GPU) acceleration. For all of our experiments, we employed one NVIDIA GeForce RTX 3060 GPU. Our GP model was trained using the RBF kernel, employing the exact marginal log likelihood as the loss function and using the Gaussian likelihood to calculate the posterior distribution. All models were initialised with a zero-prior mean. The necessary features for model training were derived from simulation data.
 Our training approach for each model integrates cross-validation \citep{hastie01statisticallearning} and early stopping \citep{Prechelt1996EarlySW}, the latter having a patience threshold set at ten epochs. The properties of each trained model are summarised in Table~\ref{tbl:training}.
\begin{table*}[h]
\caption{Training results. }
\label{tbl:training}
\centering
\begin{tabular}{ccccc}
\hline
model                                           & \multicolumn{1}{c}{best epoch} & \multicolumn{1}{c}{time/epoch (s)} & \multicolumn{1}{c}{$l$} & \multicolumn{1}{c}{$\sigma_n$} \\ \hline
$\mathcal{G}\mathcal{P}(M, \vec{r}, \vec{v})$                          & 32                             & 7.2                                & 1.6205                  & 0.1445                          \\
$\mathcal{G}\mathcal{P}(\vec{r}, \vec{v})$                        & 53                             & 6.1                                & 1.4506                  & 0.0752                         \\
$\mathcal{G}\mathcal{P}(U_{ij},K_i,K_j)$ & 12                             & 1.4                                & 0.5034                  & 0.3965                          \\
$\mathcal{G}\mathcal{P}(M)$         & 69                             & 1.0                                & 1.5523                 & 0.0040                          \\ \hline
\end{tabular}
\tablefoot{
Columns: (1) Model; (2) best epoch; (3) training time per epoch; (4) scale length, $l$; (5) noise level, $\sigma_n$.
}

\end{table*}

\begin{figure}
    \centering
    \includegraphics[width=0.5\linewidth]{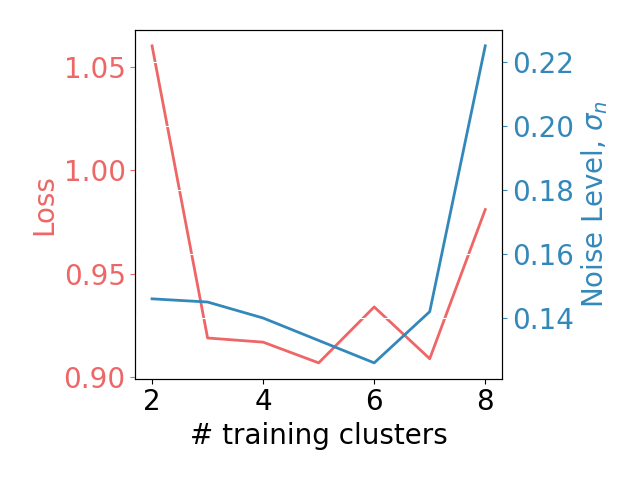}
    \caption{Training parameters (loss and noise level) with respect to the number of clusters used in the training. One validation cluster was used in the case of two or three training clusters; otherwise, we used two validation clusters. The training parameters correspond to the model's version at early stopping.}
    \label{fig:training}
\end{figure}

The models were trained on ten clusters, among which three are for validation. The training itself can be done using more or fewer clusters, depending on the context. We observed that the loss slightly increases when decreasing the number of clusters used in training (see Fig.~\ref{fig:training}), suggesting a slight decline of the model's performance. Also, we noticed a sudden increase of the loss when using eight clusters for training (and the other two for validation). This could be caused by underfitting, as suggested by the larger noise level. Therefore, we continued our experiments using a training validation ratio of seven to three. 

\subsection{Sampling}
The sampling is based on a probability density function drawn from $\mathcal{G}\mathcal{P}(M, \vec{r}, \vec{v})$  for DMCMC or $\mathcal{G}\mathcal{P}(U_{ij},K_i,K_j)$ for EMCMC. The initial state, in both cases, is initialised from a normal distribution $\mathcal{N}(\mu,\sigma)$, with $\mu$ and $\sigma$ computed over the feature distributions retrieved from the simulation data. 

The cluster size, that is, the number of stars in the cluster, $N$, is established when defining the initial state of MCMC proposals. The new candidate stars are proposed such that the state perturbations are uniformly distributed along the stars' features. This means that we do not define a maximum number of iterations, but several accepted states, $N_A$. For instance, if $N_A=mN$, new features for each star will be accepted exactly $m$ times. In this way, one can recover $m$ cluster realisations or use the first $m-1$ iterations as a burn-in strategy~\citep{doi:10.1146/annurev-statistics-031219-041300}.

\begin{figure*}
    \centering
    \includegraphics[width=\linewidth]{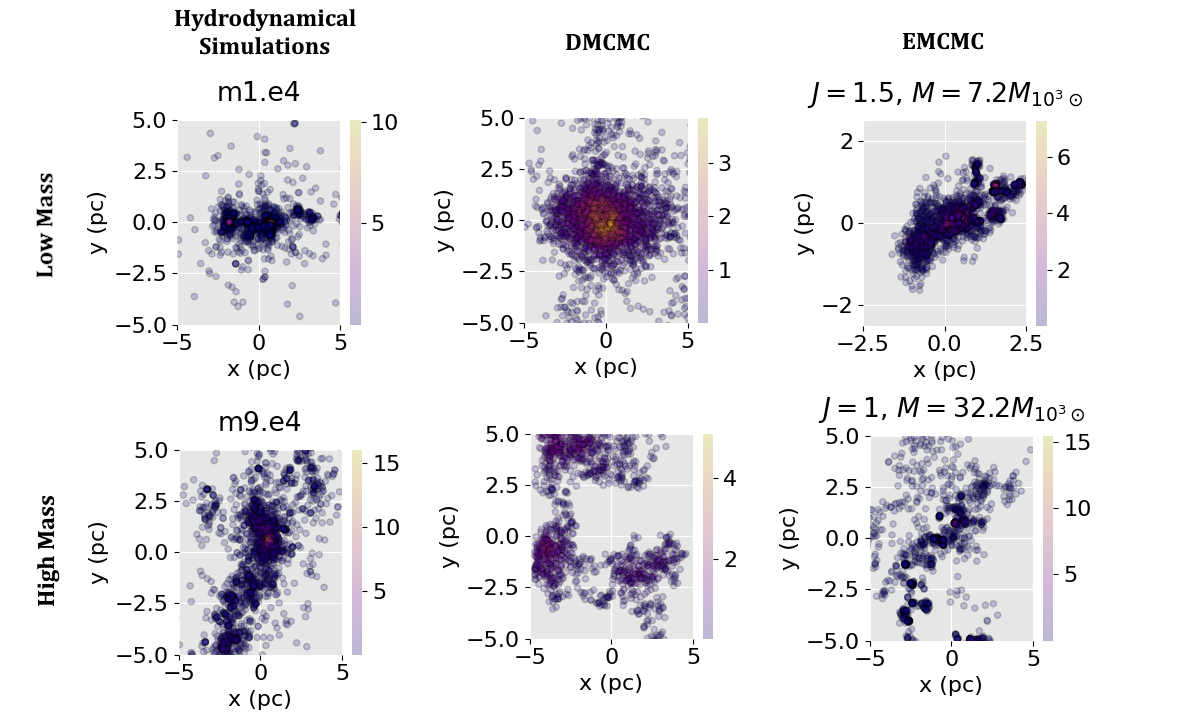}
    \caption{Two-dimensional representation of one low-mass and one high-mass cluster generated with DMCMC and EMCMC, respectively. For comparison, the corresponding projections are illustrated also for the hydrodynamically simulated clusters $m1.e4$ and $m9.e4$ from the work of \cite{Ballone_2020}. The $x$ and $y$ coordinates are measured in parsecs. For EMCMC, we show the value of the hyper-parameter $J$ (maxJump) and the total mass expressed in $M_{10^3\odot}=1000$ solar masses.  }
    \label{fig:projections}
\end{figure*}

\subsection{Generated clusters}
The newly generated clusters were evaluated based on the distributions of the inter-particle distance, velocity, and mass. In addition, we computed global physical quantities such as the virial ratio, total mass, and the number of stars that are part of a bound system. The evaluation was based on comparing our generated clusters to the hydrodynamically simulated ones retrieved from \cite{Ballone_2020}. These clusters are our baseline, and our aim was to reproduce similar features. A two-dimensional spatial projection of such clusters obtained with DMCMC and EMCMC on the $Oxy$ plane is shown in Fig.~\ref{fig:projections}.

\begin{figure*}
    \centering
        \includegraphics[width=0.3\textwidth]{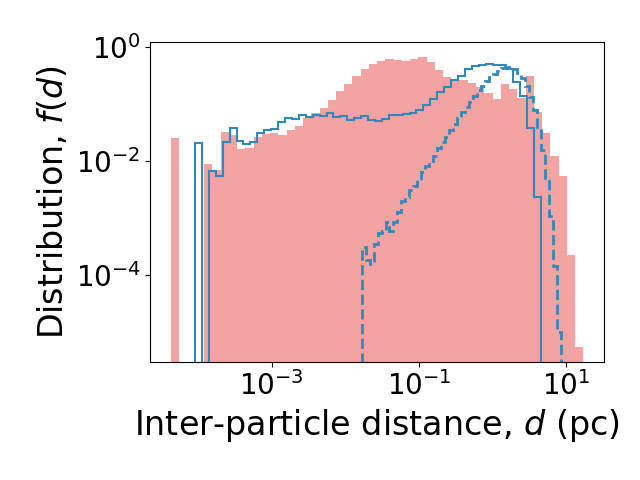}
        \includegraphics[width=0.3\textwidth]{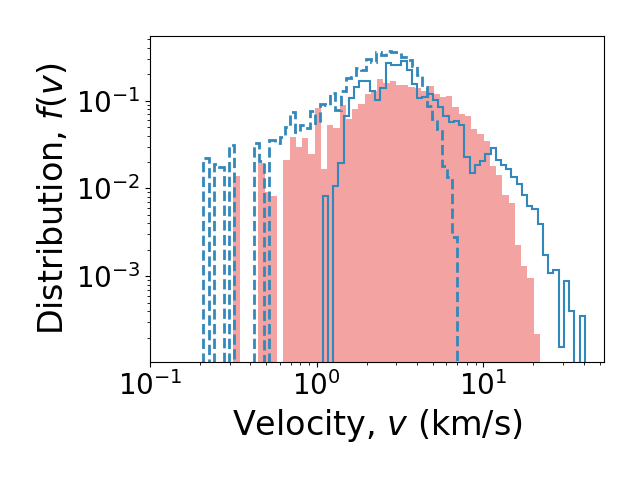}
        \includegraphics[width=0.3\textwidth]{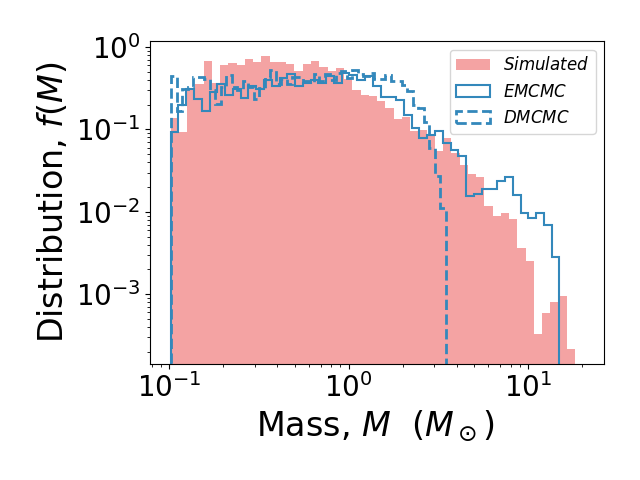}
    \caption{Distributions of inter-particle distances, velocity, and mass for low-mass clusters with 3000 stars generated using DMCMC and EMCMC methods. The mass distribution has a lower bound of $0.1$. A simulated cluster from \cite{Ballone_2020} is represented for comparison. }
    \label{fig:emcmc}
\end{figure*}

We performed a set of experiments to sample a low-mass cluster ($4000~M_\odot$) of $N=3000$ stars. The results are shown in Fig.~\ref{fig:emcmc}. Comparing to the simulation data, for the inter-particle distance distribution of DMCMC, we noticed that there is a lack of stars at low inter-particle distances smaller than $10^{-2}$ pc. This suggests a lack of close interactions between stars (i.e. binary systems). However, when analysing the total energy of the star pairs, we observed that there is a significant number of stars that are part of bound systems. For instance, the low-mass DMCMC-generated cluster that corresponds to Figure \ref{fig:projections} exhibits a number of 276 stars that are bound to other stars, whereas in the baseline training clusters there are between 400 and 800 binaries. According to the inter-particle distance distribution of the DMCMC clusters (Fig.~\ref{fig:emcmc}), the orbital separations of this binary systems are large enough that gravitational encounters will likely unbind them during a cluster's evolution. This number is thus an upper limit to stable binaries, and the actual number of stable binaries could be significantly less, given the lack of close binaries. This is not the case for the EMCMC clusters, as there are also star pairs with low inter-particle distance that are able to keep the bound system intact.

\begin{figure}
    \centering
        \includegraphics[width=0.4\textwidth]{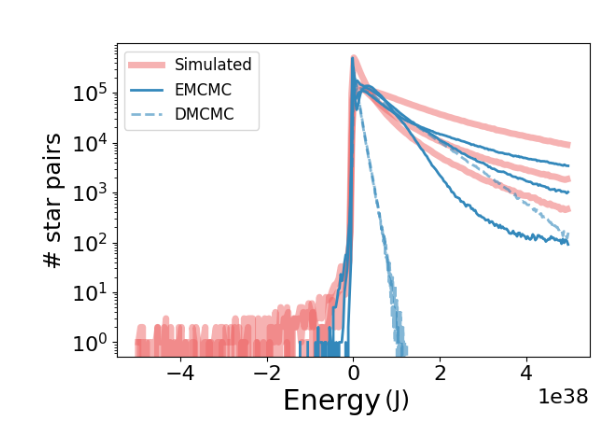}

    \caption{Histogram with the energies of each pair of stars in the cluster. The energy for each pair was calculated as the sum of kinetic energies of the stars and their mutual potential energy. Negative energies correspond to gravitationally bound pairs. The resulting energy spectra for the clusters generated with EMCMC and DMCMC are compared to the clusters obtained via hydrodynamical simulations~\citep{Ballone_2020}. }
    \label{fig:sp}
\end{figure}

On the other hand, the physics-informed algorithm excels at sampling clusters with a balanced spectrum of inter-particle distances, which direct sampling in a seven-dimensional feature space fails to achieve. We also note that the energy spectrum of DMCMC exhibits a linear trend that differs from those of the EMCMC-generated clusters or hydrodynamical simulations of clusters (see Fig.~\ref{fig:sp}). 

\begin{figure*}
    \centering
        \includegraphics[width=0.4\textwidth]{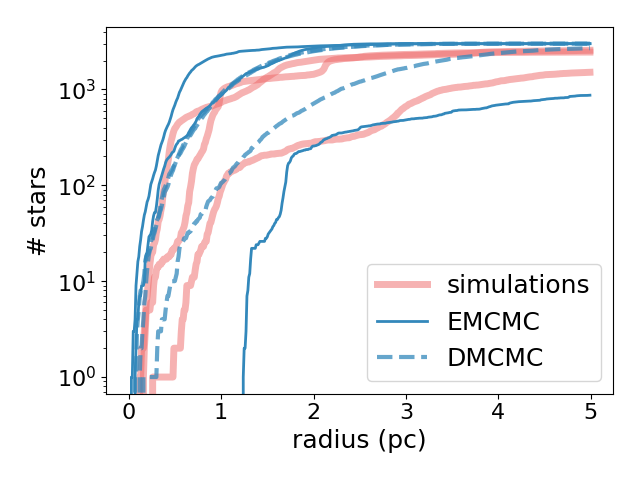}
        \includegraphics[width=0.55\textwidth]{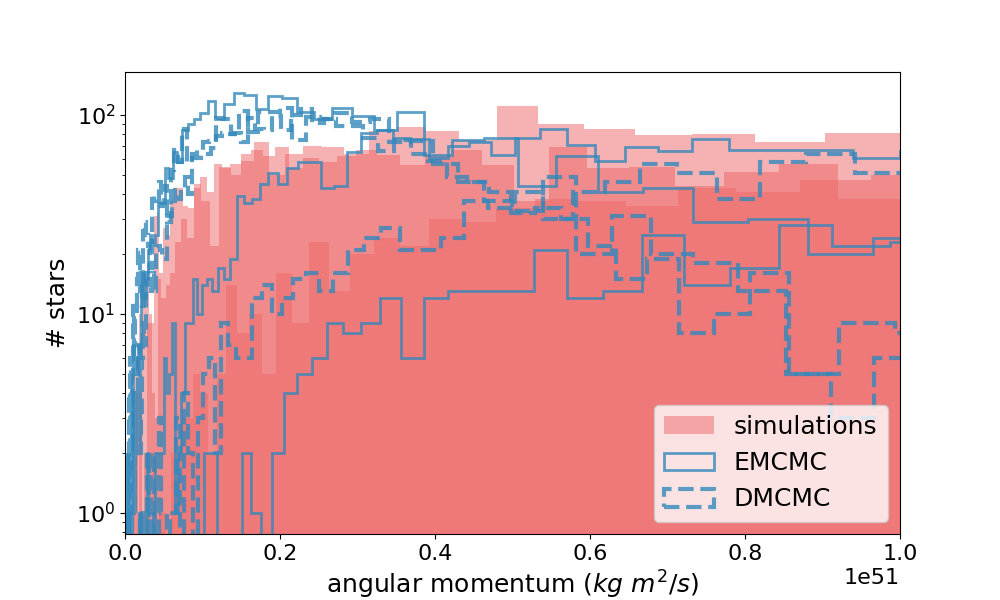}
    \caption{The distributions of key physical paramters for the new sampled clusters. The left subfigure shows the number of stars as a function of the distance to the cluster's centre of mass for a set of simulated clusters, DMCMC and EMCMC (three clusters each). On the right subfigure is plotted a histogram of the angular momentum distribution of the stars for the same sets of clusters as in the left subfigure. The angular momentum is measured in the units of the international system.}
    \label{fig:density_and_momentum}
\end{figure*}

Other relevant physical quantities are presented in Fig.~\ref{fig:density_and_momentum}. In the left subfigure, it is shown how the number of stars varies with respect to the cluster's radius. We observed that there is not any specific trend in the distribution derived for the reference clusters. We noticed sudden jumps, meaning that there are regions of higher density, but there is one exception for one of the DMCMC clusters, where the number of stars increases smoothly. On the right side, we have plotted the histogram illustrating the distribution of angular momenta resulting after computing the magnitude of angular momentum of each star as 
\begin{equation}
    |\mathbf{L}|=\sqrt{(r_yv_z - r_zv_y)^2 + (r_zv_x - r_xv_z) ^ 2 + (r_xv_y - r_y v_x)^2}.
\end{equation}
Here, we note that DMCMC clusters lack stars with high angular momentum. Probably, this happens because of the lower sampling rate of stars with high velocity magnitudes (see Fig.~\ref{fig:emcmc}).

We present a summary of additional experiments in Table \ref{tbl:metropolis}. We show the global properties (virial ratio, estimated number of binary systems, and total mass) of the six realisations obtained via DMCMC and EMCMC. We performed the experiments in order to reproduce clusters with a low, intermediate, and high total mass.

When examining virial ratios, that is, the total kinetic energy divided by half of the total binding energy, we observed that the Metropolis algorithm delivers varied dynamical states. The EMCMC approach consistently produces realistic clusters; 60\% of the clusters we sampled fall within a virial ratio between one and two, which is consistent with the training samples.

All the experiments were aimed at generating clusters containing 3000 stars, which would allow us to provide computational performance comparisons. The number $N_A=3000$ was chosen arbitrarily; one can try different cluster sizes. However, the number of stars in the clusters should be the same order of magnitude as those used in the training (2000 - 5000 stars). More massive clusters could be sampled, but the framework performance has been tested primarily for lower-mass clusters, so applying it to significantly larger clusters might necessitate retraining or fine-tuning the model to ensure that it handles the increased star count effectively. Sampling with DMCMC takes around 15 minutes for $N_A=3000$ accepted samples, while EMCMC samples 3000 new accepted states during the same time, but it needs additional time, around 10 minutes, to reconstruct the cluster from the chain. The running time will scale linearly with the number of stars, as the sampling is done sequentially. One would expect EMCMC sampling to take longer, as we are estimating the probability density for three stars at each iteration. However, the acceptance rate is also three times higher for EMCMC. The experiments were done using an NVIDIA GeForce RTX 3060 GPU and a 12th Gen Intel Core i7-12700H CPU at 4.70 GHz. The training, sampling, and evaluation routines are available online on GitHub,\footnote{\url{https://github.com/prodangp/star-clusters-gen}} and there one can find several results that we have generated and the models we trained to obtain these results.

\begin{table*}[h]
\caption{Monte Carlo sampling results using EMCMC and DMCMC algorithms. }
\label{tbl:metropolis}
\centering
\begin{tabular}{cccccccc}
\hline
 \#  & method &$M_{lower}~(M_\odot)$ & step size & acc. rate & $\alpha_{vir}$  & \# binaries & M ($10^3 M_\odot$) \\ \hline
 1& &0.1& 1.50               & 0.06    & 1.47  & 214        & 3.7       \\
 2&DMCMC&0.3&  1.50              & 0.07   & 3.52 & 75        & 10.2      \\ 
3&&0.5&  1.50                   & 0.07    & 1.26  & 261       & 24.0       \\ \hline
 4&&0.1&  0.2                 & 0.33    & 1.69 & 2089         & 7.2      \\
 5&EMCMC&0.2&  0.2                   & 0.5097    & 1.08 &   1334     &   13.5    \\
 6& &0.5& 0.4            & 0.35    & 1.21 & 1401        & 32.2       \\
\hline
\end{tabular}
\tablefoot{
The generated clusters contain 3000 stars. Columns: (1) \#, cluster index; (2) method; (3) the lower bound for the mass distribution, $M_{lower}$, in solar masses $(M_\odot)$ ; (4) the step size in the Metropolis algorithm; (5) the acceptance rate; (6) the virial ratio; (7) the number of binaries; (8) the total mass of the sampled cluster.
}

\end{table*}

\section{Conclusions}

We have introduced a learning framework to generate new realisations of positions, velocities, and masses of star clusters to be used as initial conditions in $N$-body simulations. This framework allowed us to bypass the computational bottleneck represented by hydrodynamical simulations of star formation in molecular clouds, when provided an adequate training set. Based on well understood GPs, our method yields predictable results and can be readily integrated into a simulation pipeline. 

The learning framework consists of two steps. First, a statistical model of the data is constructed through the GP, and second, new clusters are generated by sampling from the target distribution. The probability density function of the star features was employed to train the GP model. We tackled the problem in two ways: by sampling directly in the seven-dimensional feature space of the stars (DMCMC) and by sampling in a specific energy space (EMCMC). That was followed by reconstructing the cluster using the proposed algorithm.

In the pursuit of sampling clusters with realistic spatial distributions and dynamical states, the EMCMC algorithm has proven to be particularly effective, especially in addressing the crucial distribution of inter-particle distances, whereas DMCMC led to undesired distance distributions. The latter method and other conventional samplers work similar to a `black box' in this case, relying only on the correlations learned by the GP model. The main difference between the two approaches is that EMCMC incorporates physical principles. This physics-informed algorithm ensures that the generated samples are not only mathematically plausible but that they also possess physical significance.

The DMCMC experiments showed that knowing the probability density function of the mass, position, and velocity is not sufficient to fully reproduce a cluster realisation. These properties are a consequence of physical interactions, and trying to reproduce them by sampling an estimated probability density function proved to be challenging due to a lack of binary stars. In this context, the learned distribution can inform us about the possible location of a star in the feature space, but it cannot determine whether that star is part of a binary system, and it cannot reproduce its companion star. In contrast, with EMCMC, we did not change the sampling algorithm, but we sampled from a different target distribution that helps in answering both questions, namely, where the star is and whether it is a binary. The EMCMC generates clusters that not only match statistical properties but also reflect physical realities, such as binary star formation, making it a more robust and significant method for studies of stellar evolution and cluster dynamics.

\begin{acknowledgements}
This work acknowledges financial support from the European Research Council for the ERC Consolidator grant DEMOBLACK, under contract no. 770017 (PI: Mapelli). MM and ST also acknowledge financial support from the German Excellence Strategy via the Heidelberg Cluster of Excellence (EXC 2181 - 390900948) STRUCTURES.
\end{acknowledgements}

%
%
\bibliographystyle{aa}
\bibliography{aanda.bib}

\begin{thebibliography}{38}
\expandafter\ifx\csname natexlab\endcsname\relax\def\natexlab#1{#1}\fi

\bibitem[{{Aarseth}(2003)}]{2003gnbs.book.....A}
{Aarseth}, S.~J. 2003, {Gravitational N-Body Simulations}

\bibitem[{Ballone {et~al.}(2020)Ballone, Mapelli, Carlo, Torniamenti, Spera, \&
  Rastello}]{Ballone_2020}
Ballone, A., Mapelli, M., Carlo, U. N.~D., {et~al.} 2020, Monthly Notices of
  the Royal Astronomical Society, 496, 49

\bibitem[{{Ballone} {et~al.}(2020){Ballone}, {Mapelli}, {Di Carlo},
  {Torniamenti}, {Spera}, \& {Rastello}}]{2020MNRAS.496...49B}
{Ballone}, A., {Mapelli}, M., {Di Carlo}, U.~N., {et~al.} 2020, \mnras, 496, 49

\bibitem[{{Ballone} {et~al.}(2021){Ballone}, {Torniamenti}, {Mapelli}, {Di
  Carlo}, {Spera}, {Rastello}, {Gaspari}, \& {Iorio}}]{2021MNRAS.501.2920B}
{Ballone}, A., {Torniamenti}, S., {Mapelli}, M., {et~al.} 2021, \mnras, 501,
  2920

\bibitem[{Chib \& Greenberg(1995)}]{97f6cff1-d967-3115-9345-0aefb4d77f55}
Chib, S. \& Greenberg, E. 1995, The American Statistician, 49, 327

\bibitem[{{Chung} {et~al.}(2019){Chung}, {Pasquato}, {Lee}, {di Carlo}, {An},
  {Yoon}, \& {Lee}}]{2019ApJ...883L..31C}
{Chung}, C., {Pasquato}, M., {Lee}, S.-Y., {et~al.} 2019, \apjl, 883, L31

\bibitem[{Cranmer {et~al.}(2020)Cranmer, Brehmer, \&
  Louppe}]{doi:10.1073/pnas.1912789117}
Cranmer, K., Brehmer, J., \& Louppe, G. 2020, Proceedings of the National
  Academy of Sciences, 117, 30055

\bibitem[{{Dall'Amico} {et~al.}(2021){Dall'Amico}, {Mapelli}, {Di Carlo},
  {Bouffanais}, {Rastello}, {Santoliquido}, {Ballone}, \& {Arca
  Sedda}}]{2021MNRAS.508.3045D}
{Dall'Amico}, M., {Mapelli}, M., {Di Carlo}, U.~N., {et~al.} 2021, \mnras, 508,
  3045

\bibitem[{Eckhardt(1987)}]{Ulam}
Eckhardt, R. 1987, Los Alamos Science, 15, 131

\bibitem[{{Evans}(1999)}]{1999ARA&A..37..311E}
{Evans}, Neal~J., I. 1999, \araa, 37, 311

\bibitem[{Gardner {et~al.}(2018)Gardner, Pleiss, Bindel, Weinberger, \&
  Wilson}]{gpytorch}
Gardner, J.~R., Pleiss, G., Bindel, D., Weinberger, K.~Q., \& Wilson, A.~G.
  2018, GPyTorch: Blackbox Matrix-Matrix Gaussian Process Inference with GPU
  Acceleration

\bibitem[{Griffiths(2023)}]{https://doi.org/10.17863/cam.93643}
Griffiths, R.-R. 2023

\bibitem[{Hastie {et~al.}(2001)Hastie, Tibshirani, \&
  Friedman}]{hastie01statisticallearning}
Hastie, T., Tibshirani, R., \& Friedman, J. 2001, The Elements of Statistical
  Learning, Springer Series in Statistics (New York, NY, USA: Springer New York
  Inc.)

\bibitem[{{Iorio} {et~al.}(2023){Iorio}, {Mapelli}, {Costa}, {Spera},
  {Escobar}, {Sgalletta}, {Trani}, {Korb}, {Santoliquido}, {Dall'Amico},
  {Gaspari}, \& {Bressan}}]{2023MNRAS.524..426I}
{Iorio}, G., {Mapelli}, M., {Costa}, G., {et~al.} 2023, \mnras, 524, 426

\bibitem[{Jia {et~al.}(2022)Jia, Sun, Lian, \& Hou}]{jia_2022}
Jia, W., Sun, M., Lian, J., \& Hou, S. 2022, Complex \& Intelligent Systems, 8

\bibitem[{Krause {et~al.}(2020)Krause, Offner, Charbonnel, Gieles, Klessen,
  V{\'{a}}zquez-Semadeni, Ballesteros-Paredes, Girichidis, Kruijssen, Ward, \&
  Zinnecker}]{Krause_2020}
Krause, M. G.~H., Offner, S. S.~R., Charbonnel, C., {et~al.} 2020, Space
  Science Reviews, 216

\bibitem[{Krumholz {et~al.}(2019)Krumholz, McKee, \&
  Bland-Hawthorn}]{doi:10.1146/annurev-astro-091918-104430}
Krumholz, M.~R., McKee, C.~F., \& Bland-Hawthorn, J. 2019, Annual Review of
  Astronomy and Astrophysics, 57, 227

\bibitem[{{Lada} \& {Lada}(2003)}]{2003ARA&A..41...57L}
{Lada}, C.~J. \& {Lada}, E.~A. 2003, \araa, 41, 57

\bibitem[{Lueckmann {et~al.}(2021)Lueckmann, Boelts, Greenberg, Gonçalves, \&
  Macke}]{lueckmann2021benchmarking}
Lueckmann, J.-M., Boelts, J., Greenberg, D.~S., Gonçalves, P.~J., \& Macke,
  J.~H. 2021, Benchmarking Simulation-Based Inference

\bibitem[{Mack \& Rosenblatt(1979)}]{knn}
Mack, Y.~P. \& Rosenblatt, M. 1979, Journal of Multivariate Analysis, 9, 1

\bibitem[{{Malmberg} {et~al.}(2011){Malmberg}, {Davies}, \&
  {Heggie}}]{2011MNRAS.411..859M}
{Malmberg}, D., {Davies}, M.~B., \& {Heggie}, D.~C. 2011, \mnras, 411, 859

\bibitem[{{Pang} {et~al.}(2022){Pang}, {Tang}, {Li}, {Yu}, {Wang}, {Li}, {Li},
  {Wang}, {Wang}, {Zhang}, {Pasquato}, \& {Kouwenhoven}}]{2022ApJ...931..156P}
{Pang}, X., {Tang}, S.-Y., {Li}, Y., {et~al.} 2022, \apj, 931, 156

\bibitem[{{Pang} {et~al.}(2021){Pang}, {Yu}, {Tang}, {Hong}, {Yuan},
  {Pasquato}, \& {Kouwenhoven}}]{2021ApJ...923...20P}
{Pang}, X., {Yu}, Z., {Tang}, S.-Y., {et~al.} 2021, \apj, 923, 20

\bibitem[{{Parker} \& {Quanz}(2012)}]{2012MNRAS.419.2448P}
{Parker}, R.~J. \& {Quanz}, S.~P. 2012, \mnras, 419, 2448

\bibitem[{{Pfalzner} {et~al.}(2015){Pfalzner}, {Davies}, {Gounelle},
  {Johansen}, {M{\"u}nker}, {Lacerda}, {Portegies Zwart}, {Testi}, {Trieloff},
  \& {Veras}}]{2015PhyS...90f8001P}
{Pfalzner}, S., {Davies}, M.~B., {Gounelle}, M., {et~al.} 2015, \physscr, 90,
  068001

\bibitem[{{Pichardo} {et~al.}(2012){Pichardo}, {Moreno}, {Allen}, {Bedin},
  {Bellini}, \& {Pasquini}}]{2012AJ....143...73P}
{Pichardo}, B., {Moreno}, E., {Allen}, C., {et~al.} 2012, \aj, 143, 73

\bibitem[{Prechelt(1996)}]{Prechelt1996EarlySW}
Prechelt, L. 1996, in Neural Networks

\bibitem[{Rasmussen(2004)}]{Rasmussen2004}
Rasmussen, C.~E. 2004, Gaussian Processes in Machine Learning, ed. O.~Bousquet,
  U.~von Luxburg, \& G.~R{\"a}tsch (Berlin, Heidelberg: Springer Berlin
  Heidelberg), 63--71

\bibitem[{Rasmussen \& Williams(2006)}]{books/lib/RasmussenW06}
Rasmussen, C.~E. \& Williams, C. K.~I. 2006, Gaussian processes for machine
  learning., Adaptive computation and machine learning (MIT Press), I--XVIII,
  1--248

\bibitem[{{Rastello} {et~al.}(2021){Rastello}, {Mapelli}, {Di Carlo}, {Iorio},
  {Ballone}, {Giacobbo}, {Santoliquido}, \&
  {Torniamenti}}]{2021MNRAS.507.3612R}
{Rastello}, S., {Mapelli}, M., {Di Carlo}, U.~N., {et~al.} 2021, \mnras, 507,
  3612

\bibitem[{Robert \& Casella(2011)}]{Robert_2011}
Robert, C. \& Casella, G. 2011, Statistical Science, 26

\bibitem[{Roy(2020)}]{doi:10.1146/annurev-statistics-031219-041300}
Roy, V. 2020, Annual Review of Statistics and Its Application, 7, 387

\bibitem[{{Spurzem} {et~al.}(2009){Spurzem}, {Giersz}, {Heggie}, \&
  {Lin}}]{2009ApJ...697..458S}
{Spurzem}, R., {Giersz}, M., {Heggie}, D.~C., \& {Lin}, D.~N.~C. 2009, \apj,
  697, 458

\bibitem[{{Spurzem} \& {Kamlah}(2023)}]{2023LRCA....9....3S}
{Spurzem}, R. \& {Kamlah}, A. 2023, Living Reviews in Computational
  Astrophysics, 9, 3

\bibitem[{{Staff at the National Astronomy} \& {Ionosphere
  Center}(1975)}]{1975Icar...26..462.}
{Staff at the National Astronomy} \& {Ionosphere Center}. 1975, \icarus, 26,
  462

\bibitem[{{Torniamenti} {et~al.}(2021){Torniamenti}, {Ballone}, {Mapelli},
  {Gaspari}, {Di Carlo}, {Rastello}, {Giacobbo}, \&
  {Pasquato}}]{2021MNRAS.507.2253T}
{Torniamenti}, S., {Ballone}, A., {Mapelli}, M., {et~al.} 2021, \mnras, 507,
  2253

\bibitem[{{Torniamenti} {et~al.}(2022){Torniamenti}, {Pasquato}, {Di Cintio},
  {Ballone}, {Iorio}, {Artale}, \& {Mapelli}}]{Torniamenti_2021}
{Torniamenti}, S., {Pasquato}, M., {Di Cintio}, P., {et~al.} 2022, \mnras, 510,
  2097

\bibitem[{Wang(2022)}]{wang2022intuitive}
Wang, J. 2022, An Intuitive Tutorial to Gaussian Processes Regression

\end{thebibliography}

\end{document}